\documentclass[twocolumn]{aastex62}
\usepackage{amsmath}

\begin{document}

\title{{\bf High-Time-Resolution Photometry of AR Scorpii: Confirmation of the White Dwarf's Spin-Down}}

\author[0000-0002-7181-2554]{Robert A. Stiller}
\affiliation{Department of Physics, University of Notre Dame, Notre Dame, IN 46556, USA}

\author[0000-0001-7746-5795]{Colin Littlefield}
\affiliation{Department of Physics, University of Notre Dame, Notre Dame, IN 46556, USA}

\author[0000-0003-4069-2817]{Peter Garnavich}
\affiliation{Department of Physics, University of Notre Dame, Notre Dame, IN 46556, USA}

\author[0000-0003-4773-4602]{Charlotte Wood}
\affiliation{Department of Physics, University of Notre Dame, Notre Dame, IN 46556, USA}

\author{Franz-Josef Hambsch}
\affiliation{AAVSO / Vereniging Voor Sterrenkunde (VVS), Brugge, Belgium}
\affiliation{Bundesdeutsche Arbeitsgemeinschaft f\"{u}r Ver\"{a}nderliche Sterne e.V. (BAV), Berlin, Germany}
\affiliation{CBA Mol, Belgium}

\author[0000-0002-9810-0506]{Gordon Myers}
\affiliation{Center for Backyard Astrophysics, San Mateo}
\affiliation{AAVSO, 5 Inverness Way, Hillsborough, CA, USA}

\correspondingauthor{Robert Stiller}
\email{rstiller@alumni.nd.edu; clittlef@alumni.nd.edu}

\shorttitle{Confirmation of AR Sco's Spin-Down} 
\shortauthors{Stiller et al.}

\begin{abstract}

The unique binary AR~Scorpii consists of an asynchronously rotating, magnetized white dwarf (WD) that interacts with its red-dwarf companion to produce a large-amplitude, highly coherent pulsation every 1.97 minutes. Over the course of two years, we obtained thirty-nine hours of time-resolved, optical photometry of AR Sco at a typical cadence of 5~seconds to study this pulsation. We find that it undergoes significant changes across the binary orbital period and that its amplitude, phase, and waveform all vary as a function of orbital phase. We show that these variations can be explained by constructive and destructive interference between two periodic, double-peaked signals: the spin-orbit beat pulse, and a weaker WD spin pulse. Modelling of the light curve indicates that in the optical, the amplitude of the primary spin pulse is 50\%\ of the primary beat amplitude, while the secondary maxima of the beat and spin pulses have similar amplitudes. Finally, we use our timings of the beat pulses to confirm the presence of the disputed spin-down of the WD. We measure a beat-frequency derivative of $\dot{\nu} = (-5.14\pm 0.32) \times 10^{-17}$ Hz s$^{-1}$ and show that this is attributable to the spin-down of the WD. This value is approximately twice as large as the estimate from Marsh et al. (2016) but is nevertheless consistent with the constraints established in Potter \& Buckley (2018). Our precise measurement of the spin-down rate confirms that the decaying rotational energy of the magnetized white dwarf is sufficient to power the excess electromagnetic radiation emitted by the binary.

\end{abstract}


\section{Introduction}

The cataclysmic variable AR~Scorpii (AR~Sco) shows large-amplitude, highly periodic pulsations across the electromagnetic spectrum every 1.97 minutes, superimposed upon a strong waveform at the system's 3.56-h orbital period \citep{marsh16}. The system's low X-ray luminosity rules out the presence of significant accretion by the white dwarf (WD) primary from its M-dwarf companion, so neither of these signals is powered by accretion \citep{marsh16, takata}. Instead, AR~Sco has been called a white-dwarf pulsar because its pulsations consist of synchrotron radiation and are apparently powered by the spin-down of its highly magnetized ($\lesssim$~500~MG) WD, similar to neutron-star pulsars \citep{marsh16, buckley17}. 

The spin-down of the WD is a foundational conclusion from \citet{marsh16}, who calculated that the spin-down rate they detected is large enough to power AR Sco's pulsations. However, \citet{pb18} contested the significance of this spin-down after finding that the \citet{marsh16} spin-down ephemeris did not accurately predict the frequencies of the spin and orbital periods in their optical photometry. Although \citet{pb18} concluded that a linear spin ephemeris accurately described their data, they were also careful to note that their result constrained---but did not rule out---the WD spin-down. An unambiguous detection of the slowing spin rate, they wrote, would require additional observations.

AR Sco's light curve contains a number of remarkable features at different timescales. The orbital waveform is brightest at phase $\sim$0.4 and has a peak-to-peak amplitude of $\sim$1.5-2 mag in the optical, depending on the bandpass. \citet{katz17} proposes two alternative models to explain the orbital modulation and why it does not peak at superior conjunction. To provide observational constraints for these models, \citet{littlefield17} analyzed archival, ground-based photometry as well as 79 days of continuous photometry by the Kepler \textit{K2} mission. They reported that while the system's overall brightness has remained relatively stable since 2005, the orbital waveform peaked at a different phase and had a slightly lower amplitude between 2005-2007. Moreover, the \textit{K2} photometry showed aperiodic brightness fluctuations at the level of a few percent on a timescale of days \citep{littlefield17}.

\begin{figure*}
    \centering
    \includegraphics[width=\textwidth]{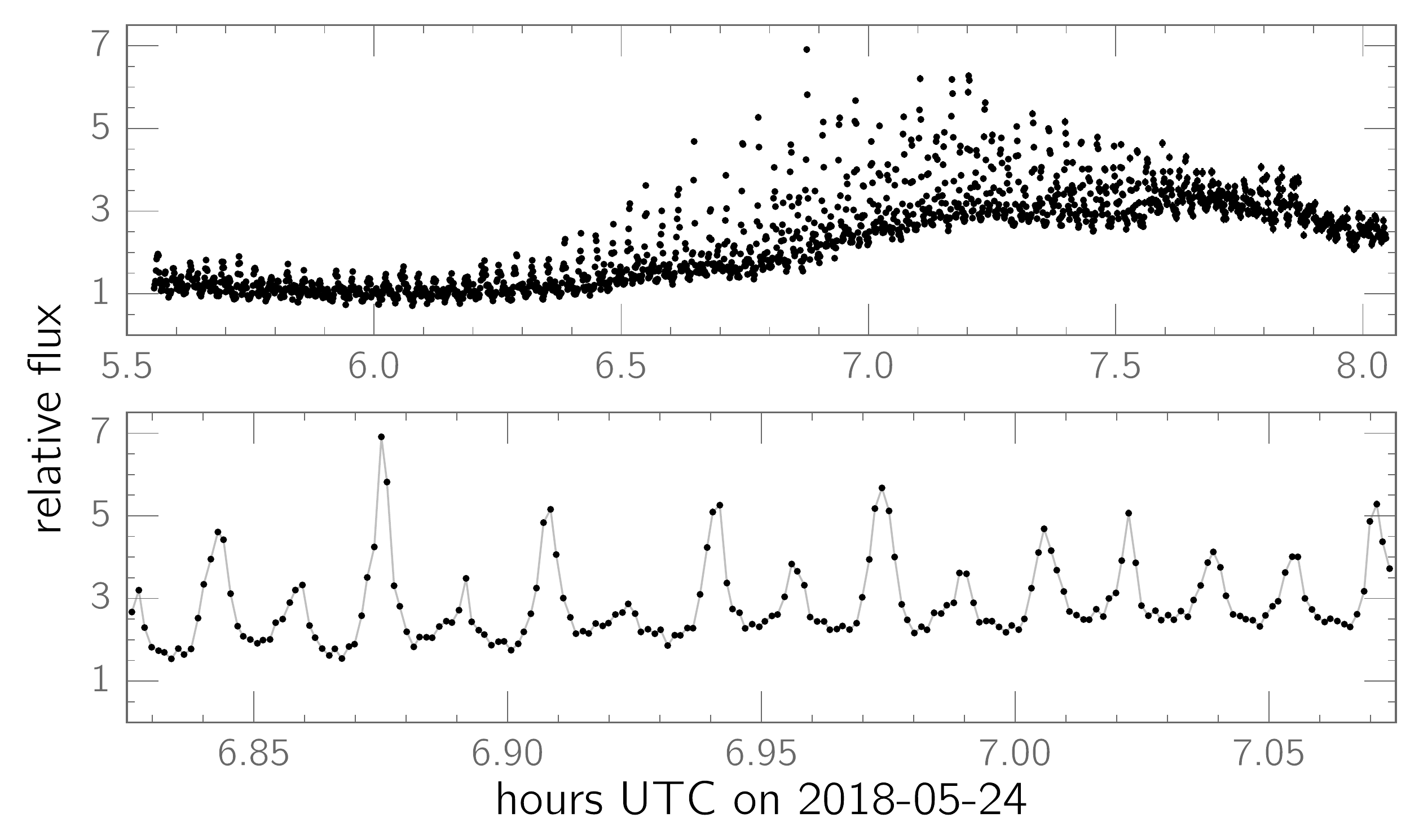}
    \caption{Sample SLKT light curve of AR Sco. The lower panel replots one section of the upper panel so that the pulsations may be seen more distinctly.}
    \label{sample_lightcurve}
\end{figure*}

The 1.97-min pulsations are arguably AR Sco's defining observational characteristic and are remarkable for their speed, amplitude (a factor of $\sim4$ in the optical), and phase coherence across a wide range of wavelengths, including radio, near-infrared, optical, ultraviolet \citep{marsh16, stanway} and even soft X-rays \citep{takata}. Their period corresponds with the beat period ({\it i.e.}, the orbital sideband) between the binary orbital period and 1.95-min WD spin period, and they are thought to originate on the inner hemisphere of the M5-class companion star \citep{marsh16, takata}. \citet{geng16} proposes that the WD's magnetic axis is inclined with respect to its rotational axis and that the pulses are caused by the interaction of the WD's magnetosphere with the secondary's wind.

The 1.95-min spin period of the white dwarf in AR~Sco is extremely short when compared with the system's orbit. White dwarfs are not born spinning so rapidly, and it is thought that a phase of high accretion powered AR~Sco's rapid spin-up, followed by the current epoch of little or no mass transfer \citep{marsh16, buckley17}.

Here, we present high-cadence photometry with the twin objectives of (1) searching for a spin-down and (2) disentangling the spin and beat periods.

\section{Data}

\subsection{SLKT observations}

We obtained 39 hours of high-time-resolution photometry of AR Sco using the 80-cm Sarah L. Krizmanich Telescope (SLKT) and an unfiltered Santa Barbara Instrument Group STL-1001 CCD camera at the University of Notre Dame in 2016, 2017, and 2018. As indicated in Table~\ref{log}, which lists details of each time series, the exposure time was 2~s, and factoring in the overhead between images, the typical cadence was 5-6~s, and each time series usually spanned 1-3~h. Fig.~\ref{sample_lightcurve} plots a representative, 2.5-hour-long light curve and zooms in on one segment during which the pulsations were especially prominent.

\begin{figure}
    \centering
    \includegraphics[width=\columnwidth]{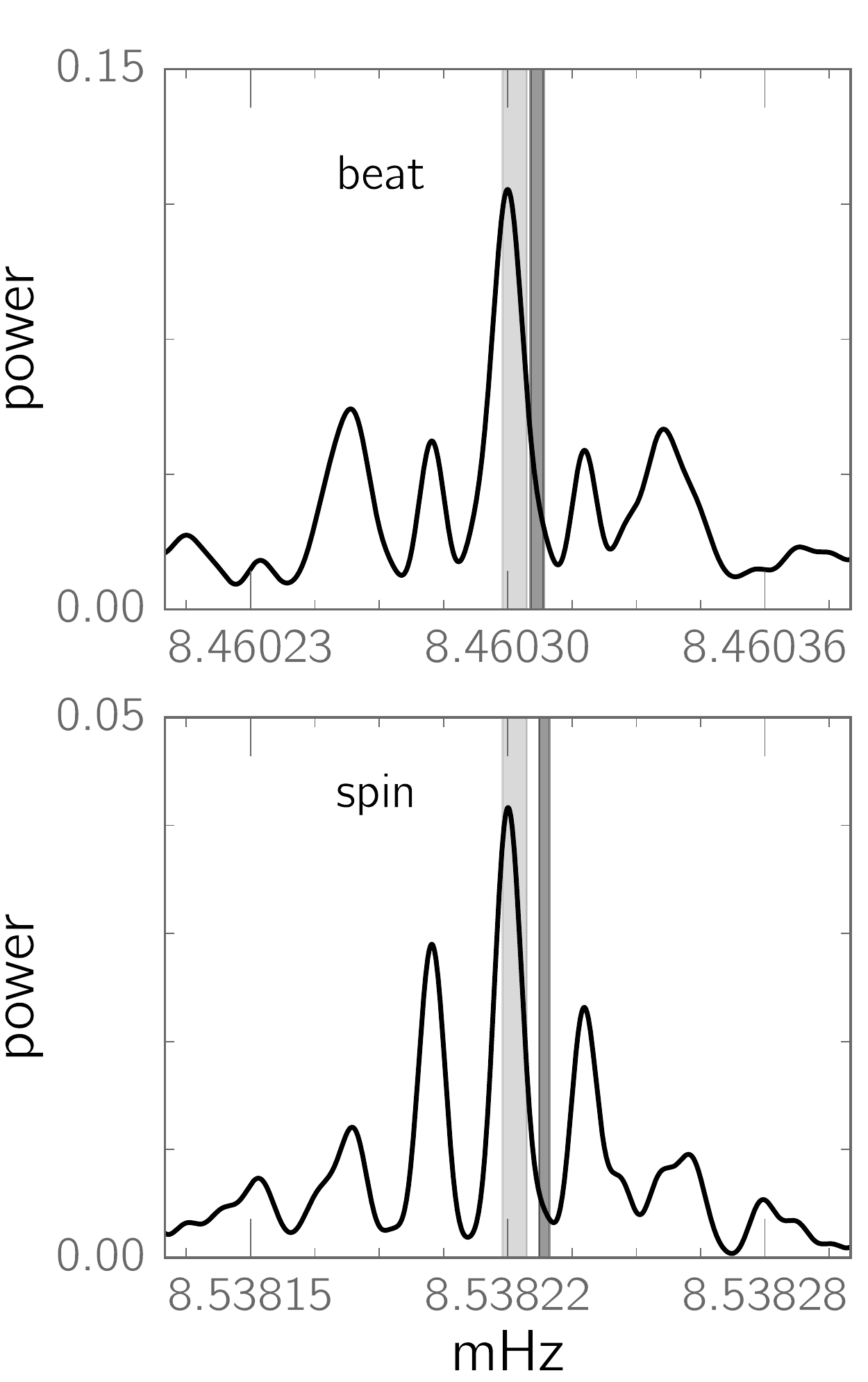}
    \caption{Lomb-Scargle power spectra of the SLKT data, focused on the beat and spin frequencies. The Lomb-Scargle model for both power spectra used two harmonic terms. The light contour gives the $1\sigma$ confidence interval from \citet{pb18}, while the dark contour is the projected confidence interval from the spin-down ephemeris in \citet{marsh16}.}
    \label{power_spectra}
\end{figure}

\begin{figure}
\includegraphics[width=\columnwidth]{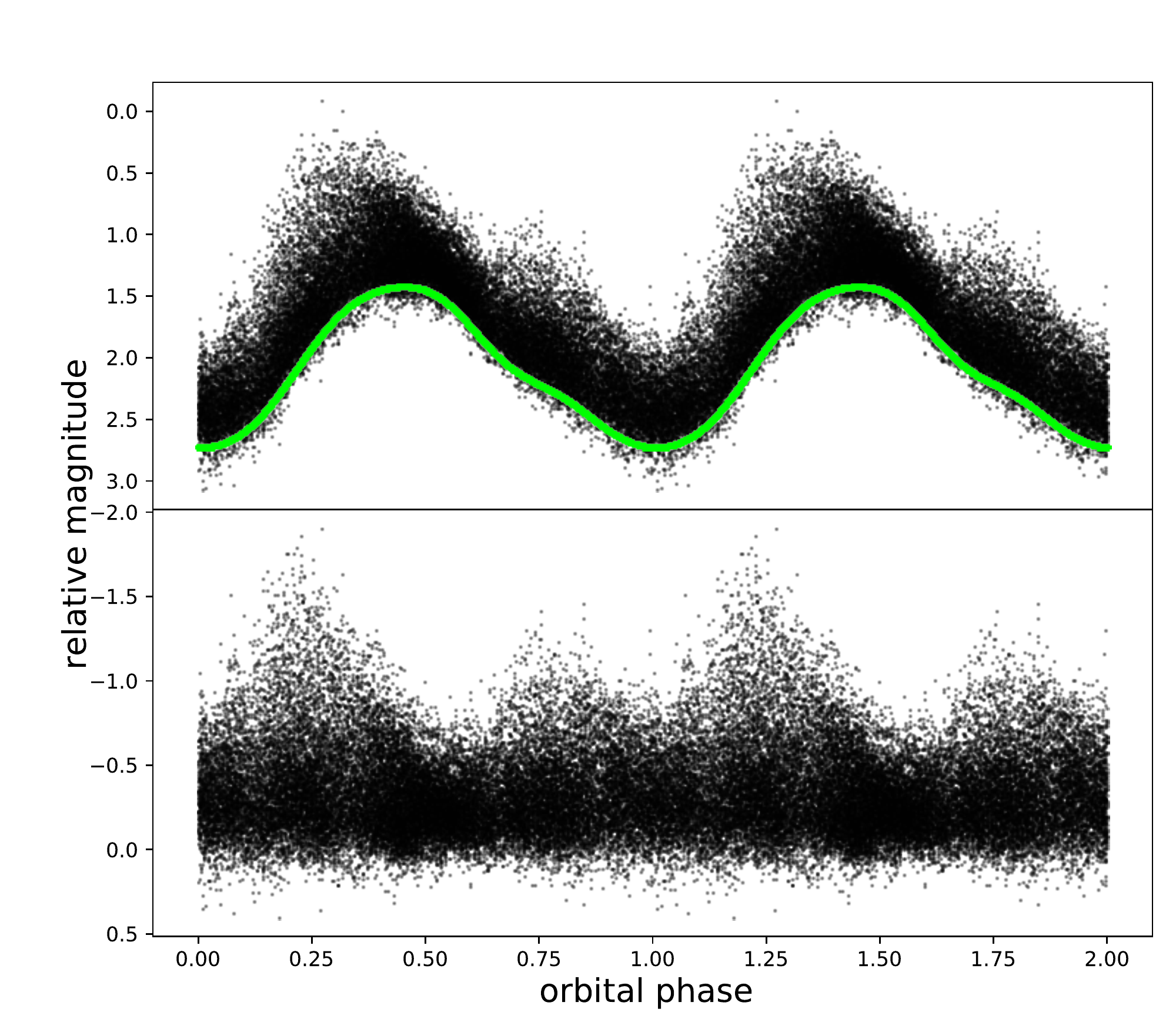}
\caption{{\bf Top:} The light curve of AR~Sco phased on the system's orbital period. The scatter results from the $\sim$2-min periodic flashes that are incoherent at this phasing. The green line is a Fourier series with 3  harmonics that traces the base of these pulsations, and it represents the orbital modulation in the absence of pulsations.  {\bf Bottom:} The residuals after subtraction of the orbital modulation. The amplitude of the pulsed variation peaks at orbital phases $\sim$0.25 and $\sim$0.75.}
\label{orbital_phaseplot}
\end{figure}

The system's optical pulsations are extremely fast and well-defined, so CCD data must be well-timed in order to be useful. Consequently, we synchronized the clock of the CCD control computer to an atomic clock prior to the start of each time series. Additionally, we measured the shutter lag of the detector (the time offset between the shutter actuation and the timestamp recorded in the FITS header), found it to be stable from night-to-night, and applied an appropriate correction to the image timestamps. Finally, we applied a BJD$_{TDB}$ correction to all observations using routines in Astropy \citep{astropy}.

We used aperture photometry to extract AR Sco's light curve, but there were no optimal comparison stars within the field of view. According to APASS photometry, each of the field stars is quite red, probably because there is a dark nebula associated with the $\rho$ Ophiuchi complex along the line of sight to AR Sco. Faced with this paucity of choices, we selected UCAC4 336-082341 ($\alpha_{2000}$ = 16h 22m 06.481s, $\delta_{2000} = -22^{\circ}$ 53' 27.84'', $g'-r' = 2.17$) as our comparison star. Since the spectral properties of such a reddened star are likely a poor match for those of AR Sco, any attempt to infer standard $V$ magnitudes from the unfiltered photometry would probably suffer from serious systematic errors. Consequently, we do not place our photometry on a standard magnitude scale.

\subsection{AAVSO observations}

As indicated in Table~\ref{aavso_log}, coauthors FJH and GM observed AR Sco and submitted their observations to the AAVSO International Database\footnote{https://www.aavso.org} under AAVSO observer codes HMB and MGW, respectively. Their observations had cadences that ranged from 14-35~s.

\begin{figure}[h]
\includegraphics[width=\columnwidth]{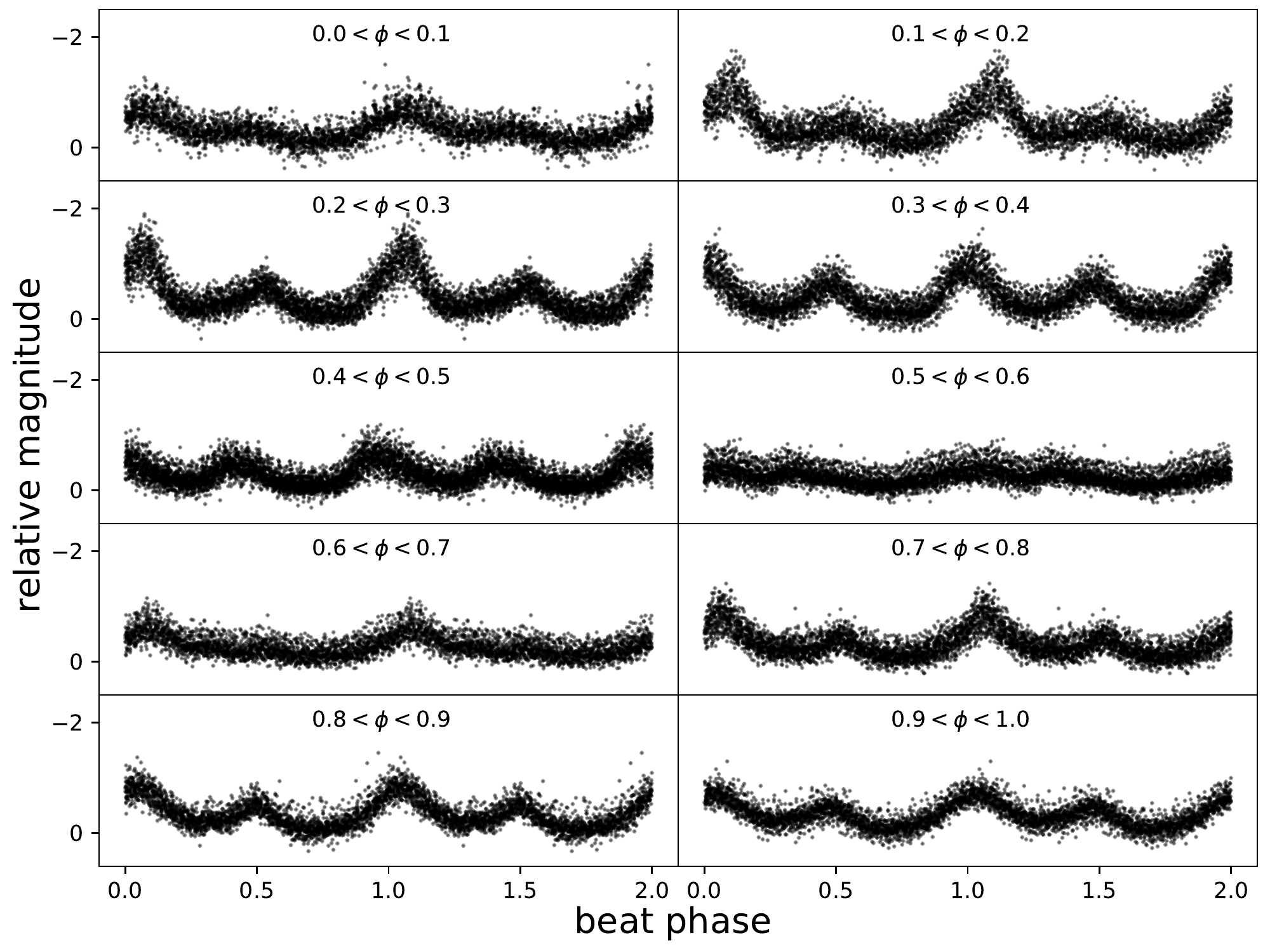}
\caption{The beat pulse shape as a function of orbital phase. After subtracting the orbital modulation, the light curve was divided into ten orbital phase bins, and each bin is phased on the beat period. Two beat cycles are shown in each panel. Two unequal amplitude peaks are seen each beat cycle suggesting that both poles of the WD magnetic field are interacting with the secondary star. The amplitude and pulse shape are seen to vary with orbital phase. The pulse shape is asymmetric during the first half of an orbit, resulting in its centroid shifting in phase by over 10~s over an orbit.}
\label{beat_bins}
\end{figure}

\begin{table}
\centering
\caption{Log of SLKT observations}
\label{log}
\begin{tabular}{ccccc}
\hline
UT Start Date & Length (hr) & Cadence (s) \\
\hline
2016-07-28 & 2.0 & 6 \\
2016-08-03 & 1.6 & 5 \\
2016-08-22 & 0.9 & 5 \\
2016-08-23 & 1.1 & 5 \\
2016-09-01 & 1.0 & 5 \\
2016-09-02 & 0.9 & 5 \\
2016-09-03 & 0.7 & 4 \\
2016-09-04 & 0.7 & 4 \\
2017-04-23 & 2.9 & 5 \\
2017-05-07 & 2.3 & 5 \\
2017-05-08 & 1.8 & 5 \\
2017-05-15 & 1.0 & 5 \\
2017-05-17 & 1.9 & 5 \\
2017-06-01 & 3.1 & 5 \\
2017-06-02 & 2.1 & 5 \\
2017-06-03 & 0.8 & 4 \\
2017-07-07 & 1.6 & 5 \\
2017-08-12 & 0.9 & 5 \\
2018-02-26 & 1.4 & 5 \\
2018-03-18 & 1.0 & 5 \\
2018-03-25 & 1.7 & 7 \\
2018-03-26 & 2.8 & 5 \\
2018-04-18 & 1.3 & 5 \\
2018-04-21 & 0.8 & 5 \\
2018-05-24 & 2.5 & 5 \\
\hline
\end{tabular}
\end{table}

\begin{table}
\centering
\caption{Log of AAVSO observations. The ``Obs.'' column contains the AAVSO code of the observer.}
\label{aavso_log}
\begin{tabular}{cccc}
\hline
Observer & UT Start Date & Length (hr) & Cadence (s) \\
\hline
HMB & 2015-07-24 & 1.7 & 35 \\
HMB & 2015-07-25 & 4.2 & 35 \\
HMB & 2015-07-26 & 4.3 & 35 \\
HMB & 2015-07-27 & 4.2 & 35 \\
HMB & 2015-08-07 & 2.7 & 35 \\
HMB & 2015-08-08 & 2.3 & 35 \\
HMB & 2015-08-10 & 2.1 & 35 \\
HMB & 2016-04-28 & 4.8 & 34 \\
HMB & 2016-04-29 & 5.7 & 24 \\
HMB & 2016-04-30 & 5.8 & 24 \\
HMB & 2016-05-01 & 5.9 & 24 \\
HMB & 2016-05-02 & 5.2 & 24 \\
HMB & 2016-05-03 & 5.7 & 24 \\
HMB & 2016-05-04 & 5.7 & 24 \\
HMB & 2016-05-05 & 5.8 & 14 \\
HMB & 2016-07-31 & 2.6 & 35 \\
MGW & 2016-08-05 & 4.1 & 24 \\
MGW & 2016-08-16 & 2.3 & 19 \\
MGW & 2016-09-04 & 3.7 & 23 \\
HMB & 2016-09-09 & 2.7 & 35 \\
HMB & 2016-09-10 & 2.6 & 35 \\
\hline
\end{tabular}
\end{table}

\section{Analysis}
\label{sec:analysis}

Lomb-Scargle power spectra of the SLKT data, presented in Fig.~\ref{power_spectra}, show that the measured spin and beat frequencies agree with those reported in \citet{pb18}. However, the predicted spin and beat frequencies from the \citet{marsh16} spin-down ephemeris are in poor agreement with the peaks observed in our data, confirming the result in \citet{pb18}. We explore this issue in depth in Sec.~\ref{sec:spin_down}.

After eliminating a small number of SLKT observations with a signal-to-noise ratio of less than 5, we display our photometry in Fig.~\ref{orbital_phaseplot}. The data have been phased to the orbital period based on the ephemeris in \citet{marsh16}. The zero point of the orbital phase is defined as the moment that the red secondary star is at inferior conjunction. The overall light curve shape is similar to the slower cadence data analyzed by \citet{littlefield17}. The light curve is asymmetric, with a quick rise to maximum brightness at an orbital phase of 0.4 and a second peak at an orbital phase of $\sim$0.75.

\begin{figure}
\includegraphics[width=\columnwidth]{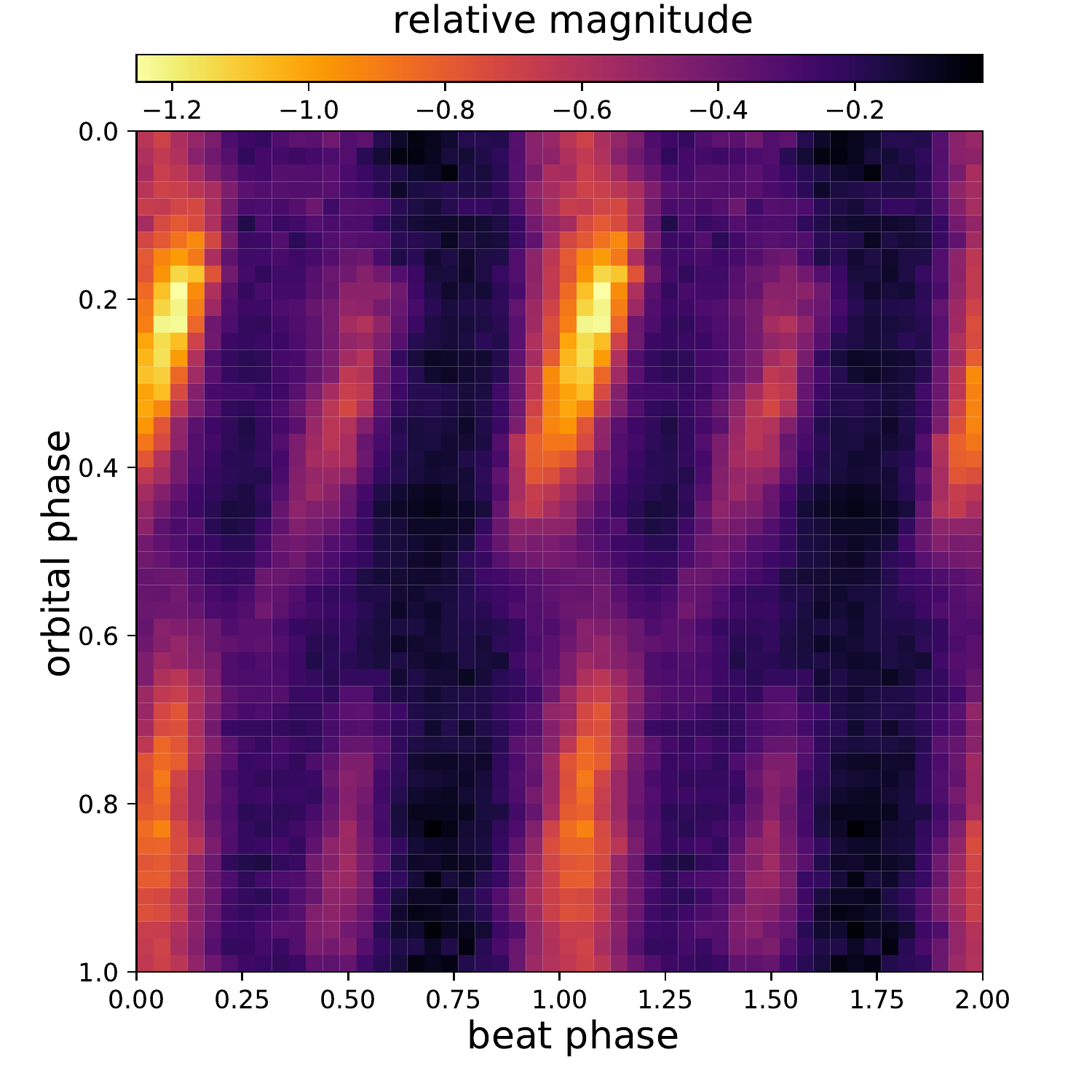}
\caption{The AR~Sco light curve phased on the beat period as it varies over a binary orbit. The orbital waveform has been removed. The major beat pulse peaks at orbital phase $\sim$0.25, then shifts to earlier beat phases before reaching a minimum at orbital phase $\sim$0.55.}
\label{drifting_beat}
\end{figure}

We assume that the observed light curve can be represented as the sum of a slowly varying orbital modulation and the high-frequency, pulsed emission. To remove the variations associated with the orbit, we divided the orbit into phase bins and identified the faintest 5\%\ of the points in each bin. Because contamination by the pulsed emission inflates the amplitude of the orbital waveform in a simple Lomb-Scargle power spectrum, the faintest observations in these bins more accurately describe the underlying orbital modulation, and we selected the exact threshold from trial-and-error. We then represented these points with a Fourier series so that we could predict the strength of the orbital modulation as a function of orbital phase. As seen in Fig.~\ref{orbital_phaseplot}, the orbital variation is moderately asymmetric, with the peak brightness just before phase 0.5 and the minimum at phase 0.0.

\begin{figure}
\includegraphics[width=\columnwidth]{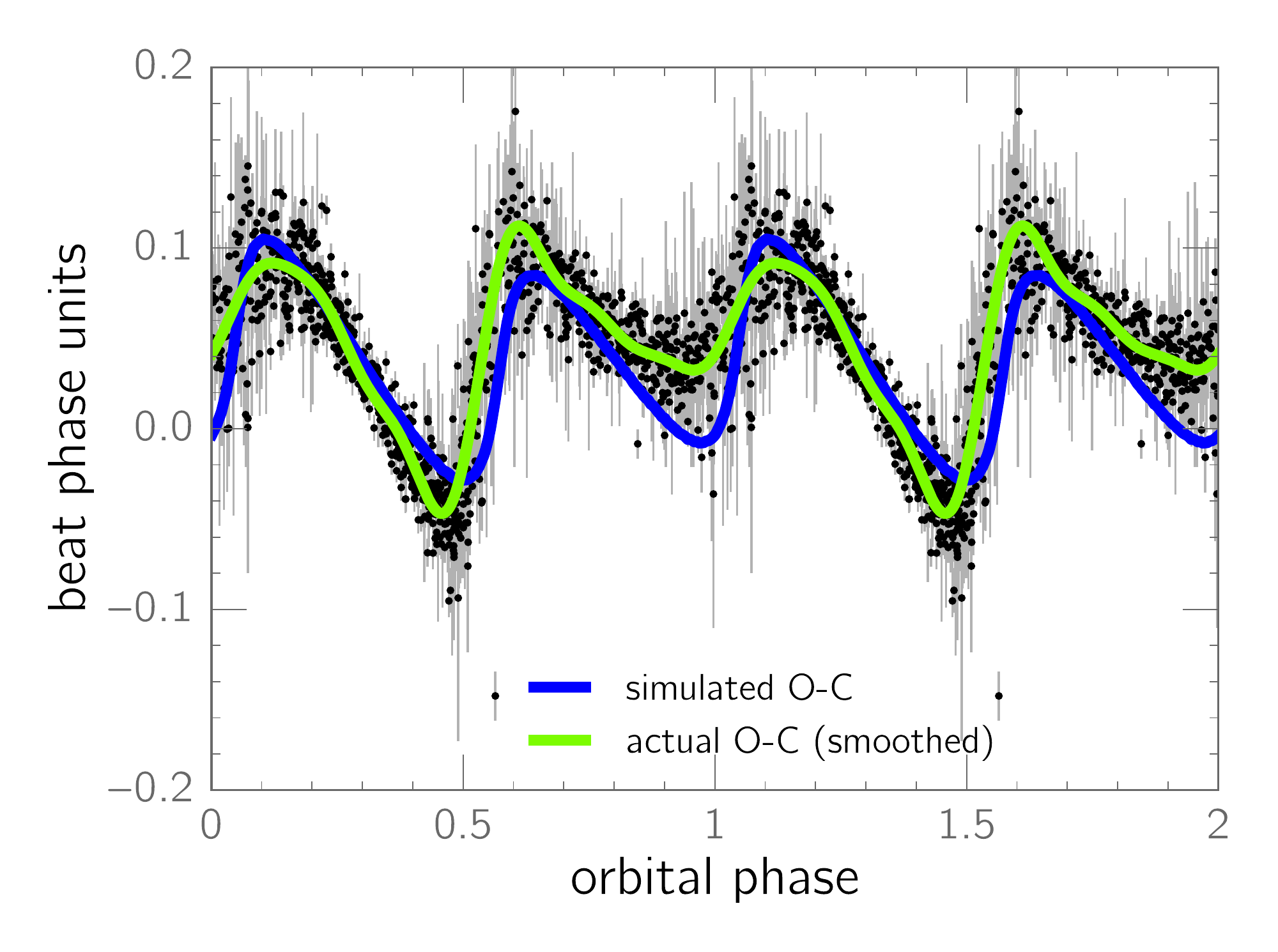}
\caption{An O$-$C diagram of the beat pulses reveals an obvious orbital-phase dependence. The increased scatter in the residuals near orbital phase 0.5 corresponds with a dropoff in the amplitude of the beat pulse. Decreased SNR near orbital phase 0.0 contributes to the noisy timings near that phase.}
\label{O-C}
\end{figure}

Subtracting the orbital-modulation function from the phased orbital light curve yields the pulsed light curve as a function of orbital phase (bottom panel in Fig.~\ref{orbital_phaseplot}). The pulse amplitude displays a strong orbital-phase dependence, peaking at orbital phase $\sim$0.25, reaching minimum amplitude near phase $\sim$0.6, and rebounding around phase 0.75.

\begin{figure*}
\includegraphics[width=\textwidth]{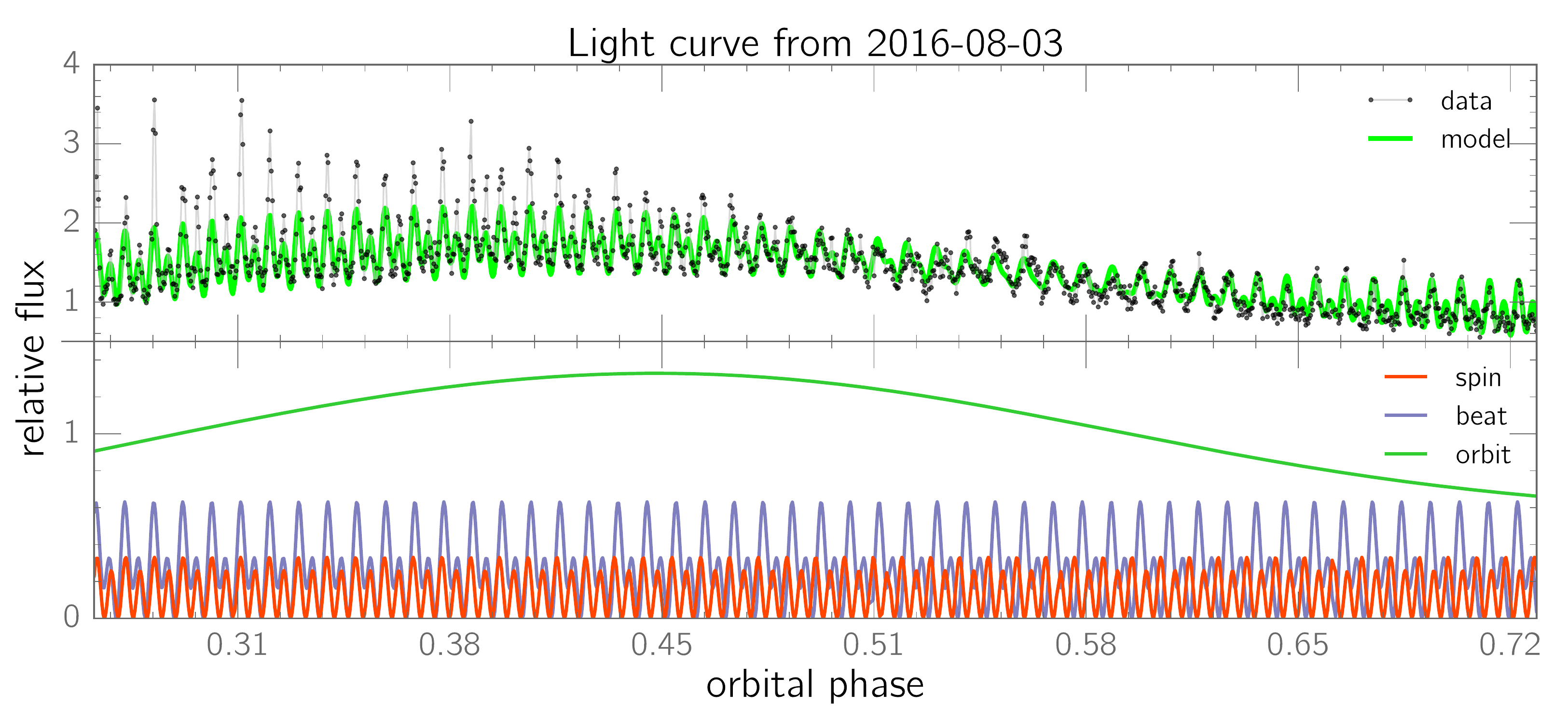}
\caption{{\bf Top:} An example of our data covering nearly half a binary orbit (points connected by a gray line). The thick line shows the light curve model combining a double peaked beat pulse with a double peaked spin pulse.  {\bf Bottom:} The three components used to build the model are an orbital modulation, the beat pulse, and the spin pulse. The spin and beat components each have two harmonic terms. The superposition of the spin and beat pulses causes the combined pulse profile to vary across the orbital period. This simple model accounts for the pulse amplitude and phase variations of over the orbit, but tends to underestimate the heights of the peaks over the brightest portion of the light curve. }
\label{model}
\end{figure*}

To more closely investigate the changes in the pulse amplitude and morphology, we split the data into ten equally-sized, non-overlapping orbital bins and phased each bin to the beat period. The resulting plots (Fig.~\ref{beat_bins}) show that the beat pulse has two unequal maxima per cycle, as has been noted by \citet{marsh16}. Our beat phaseplots show that the shape, amplitude, and phase of both maxima vary as a function of orbital phase. At orbital phase 0.0, the major pulse is broad and symmetric, but between orbital phases 0.1-0.3, its shape changes, with the rise to beat maximum becoming longer than the decline. This is also the orbital phase when the major pulse has the largest amplitude. Around orbital phase 0.35, the major peak is again symmetric but soon skews in the other direction. Eventually, the major peak nearly disappears around phase 0.55. For the second half of the orbit, the major pulse remains symmetrical, but is weaker than in the first half of the orbit. The minor pulse reaches its largest amplitude between orbital phases 0.3 and 0.4, significantly later than phase 0.25 for the major pulse.

The beat pulse undergoes a significant phase shift, depending on the orbital phase. This is most easily seen in a heatmap of the pulse brightness versus both the orbital and beat phase (see Fig.~\ref{drifting_beat}). This figure clearly shows that between orbital phases 0.1 and 0.5, the major peak shifts by 10\%\ in beat phase. The shift in the beat pulse over the second half of the orbit is smaller than the first half, but the pulse is generally seen to be broader and fainter. \citet{takata} reported similar findings from their analysis of 39~ks of data obtained on 2016 September 19 with the XMM Newton satellite's Optical/UV Monitor Telescope, and our results imply that this behavior is stable on timescales of years.

The amplitude and phase shifts of the beat pulse are consistent with the addition of second periodic signal with a slightly different frequency \citep[e.g., as observed in FO Aqr;][]{om89}. For example, if the spin pulse were to be isolated and plotted in the heat map in Fig.~\ref{drifting_beat}, it would run diagonally since the figure phases the data to the beat period. To explore this possibility, we modeled the full light curve as the superposition of three periodic signals: the spin, beat, and orbital periods (similar to the light curve model for FO~Aqr in its low state \citep{littlefield16}). We found the best-fit trigonometric function at each of those frequencies by a simple least-squares fit, with each term consisting of three harmonics. We also attempted to fit additional frequencies detected in the power spectrum, but adding these terms did not significantly improve the quality of the fit.

The results of this fit (Fig.~\ref{model}) reveal that the beat and spin models are both double-peaked, with the beat pulse having two unequal maxima. The maxima of the spin pulse, by contrast, are roughly equal. Our model predicts that the amplitude of the spin pulse is about 50\%\ of the amplitude of the beat pulse, while the minor pulses are comparable in brightness. However, one limitation of our model is that it underestimates the amplitude of the highest-amplitude beat pulses. This might be a consequence of the fact that the model does not account for the changing visibility of the secondary's inner hemisphere across the orbital period.

As a test of this model, we computed the O$-$C for the blended spin-beat pulse, and as shown in Fig.~\ref{O-C}, it offers a reasonably accurate prediction of the actual O$-$C. The simulated O$-$C includes a correction for light-travel delays caused by the secondary's orbital motion. The light-travel delay for pulsations originating from material orbiting in the plane of the binary is given by \begin{equation} \Delta t = -\frac{d}{c}\sin(i)\cos(2\pi[\phi_{orb}-\phi_0]),\label{light-travel}\end{equation} where $d$ is the distance of the emission from the binary center of mass, $i$ is the orbital inclination, $c$ is the speed of light, and $\phi_0$ is orbital phase of inferior conjunction for the emitting material. Given the requirement that the emission originate on the donor \citep{marsh16}, the smallest-possible light-travel delay would occur for emission arising at the first Lagrangian point (L$_1$). If we adopt M$_{1}$ = 0.8 M$_{\odot}$ and M$_{2}$ = 0.3 M$_{\odot}$, as did \citet{marsh16}, and assume an orbital inclination of $i = 60^{\circ}$, then the semi-amplitude of the light-travel delay would be 0.6~s. 

Although the effect of light-travel delays is relatively minor at our $\sim$5-s cadence, a higher time resolution might potentially be able to discern light-travel delays from emission from different regions of the secondary (\textit{i.e.}, across a range of values of $d$ and $\phi_{0}$).

Although there might be a light-travel delay associated with the spin pulse, the region in which the optical spin pulse is generated is unknown, making it impossible to calculate its light-travel delay. \citet{buckley17} provide evidence that it might originate in a ``striped wind" outside the light cylinder of the magnetosphere, which is almost an order of magnitude larger than the binary orbital separation.

\section{Spin-down ephemeris}\label{sec:spin_down}

\subsection{Measuring the spin-down}

\citet{marsh16} detected a significant spin-frequency derivative of $\dot{\nu} = -(2.86\pm0.36)\times10^{-17}$ Hz s$^{-1}$, implying a spin-down of sufficient magnitude to power the optical pulsations. However, \citet{pb18} found that the \citet{marsh16} spin-down ephemeris did not accurately predict the frequencies observed in their power spectra and concluded their photometry was consistent with a constant spin period of 0.008538220(3)~Hz, where the number in parentheses is the uncertainty on the final digit. While \citet{pb18} showed that the spin-down ephemeris from \citet{marsh16} was inaccurate, they noted that their results could still be reconciled with a nonzero $\dot{\nu}$ and that a longer baseline of observations was necessary to investigate this possibility.

Our results in Sec.~\ref{sec:analysis} show that the beat pulse is more readily measured than the lower-amplitude spin pulse, so we use O$-$C measurements of the beat pulse to search for a change in the spin period. Assuming that any change in the orbital period is small over the baseline of observations, the derivative in the beat frequency will be a direct measure of the derivative of the spin frequency.

\begin{figure}
\includegraphics[width=\columnwidth]{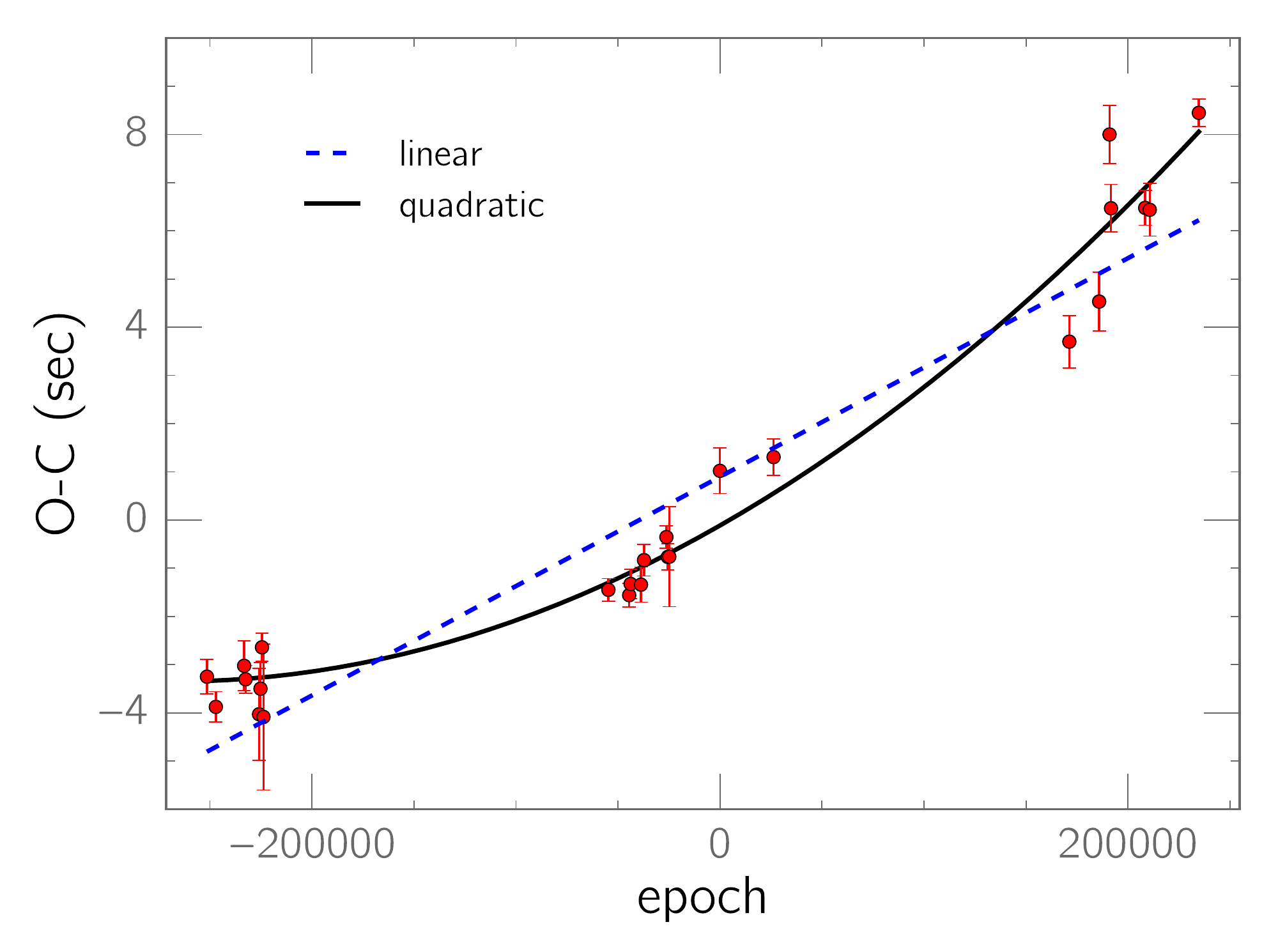}
\caption{An O$-$C diagram of the SLKT beat-pulse timings using the \citet{pb18} beat period. Each point is the mean residual from a given observing run, and each errorbar is the standard error of that mean. A quadratic fit to the residuals yields a noticeably improved fit to the data compared to a simple linear fit, indicating that the measured O$-$C values are the result of a period derivative and not simply an inaccurate period.}
\label{O-C_PB18}
\end{figure}

We measured 1,077 beat-pulse timings\footnote{These timings are available as an online table, and in Table~\ref{pulse_timings}, we provide a sample of them to illustrate the format of the data.} from our dataset by fitting a Gaussian to each well-observed beat pulse. The timings, which span two years and three observing seasons, have a sufficiently long baseline to search for $\dot{\nu}$. We applied a correction to all pulse timings to compensate for the orbital-phase dependence of their arrival times, using the empirical fit from Fig.~\ref{O-C}. We then tested the \citet{pb18} linear ephemeris by computing an O$-$C diagram, using the beat period implied by their spin period. To improve the signal-to-noise ratio, we averaged the O$-$C residuals for each night and used the standard error of their mean as the 1$\sigma$ uncertainty for each night. The residuals in this plot (Fig.~\ref{O-C_PB18}) show a rising trend, suggesting that the beat period inferred from \citet{pb18} is not a good match to our data. Further, there appears a significant curvature in the $O-C$ measurements consistent with the presence of a period derivative.

\begin{table}
\centering
\caption{Beat-pulse timings. The full table is available online as a machine-readable table.}
\label{pulse_timings}
\begin{tabular}{cccc}
\hline
Epoch\tablenotemark{a} & T$_{ max }$ [BJD]\tablenotemark{b} & T$_{ max,corr}$[BJD]\tablenotemark{c} & $\pm$ [d]\ \\
\hline
0 & 2457941.668881 & 2457941.668860 & 0.000008 \\
1 & 2457941.670226 & 2457941.670210 & 0.000009 \\
2 & 2457941.671584 & 2457941.671572 & 0.000011 \\
3 & 2457941.672936 & 2457941.672930 & 0.000008 \\
4 & 2457941.674330 & 2457941.674329 & 0.000010 \\
5 & 2457941.675686 & 2457941.675691 & 0.000011 \\
6 & 2457941.677076 & 2457941.677088 & 0.000008 \\
7 & 2457941.678407 & 2457941.678427 & 0.000010 \\
8 & 2457941.679758 & 2457941.679786 & 0.000010 \\
9 & 2457941.681116 & 2457941.681153 & 0.000012 \\
\hline
\end{tabular}
\raggedright
\tablenotetext{a}{Relative to Eq.~\ref{ephem}.}
\tablenotetext{b}{Raw pulse timings, uncorrected for orbital-phase dependence of arrival times.}
\tablenotetext{c}{Pulse timings corrected for orbital phase.}
\end{table}

\begin{figure}
\includegraphics[width=\columnwidth]{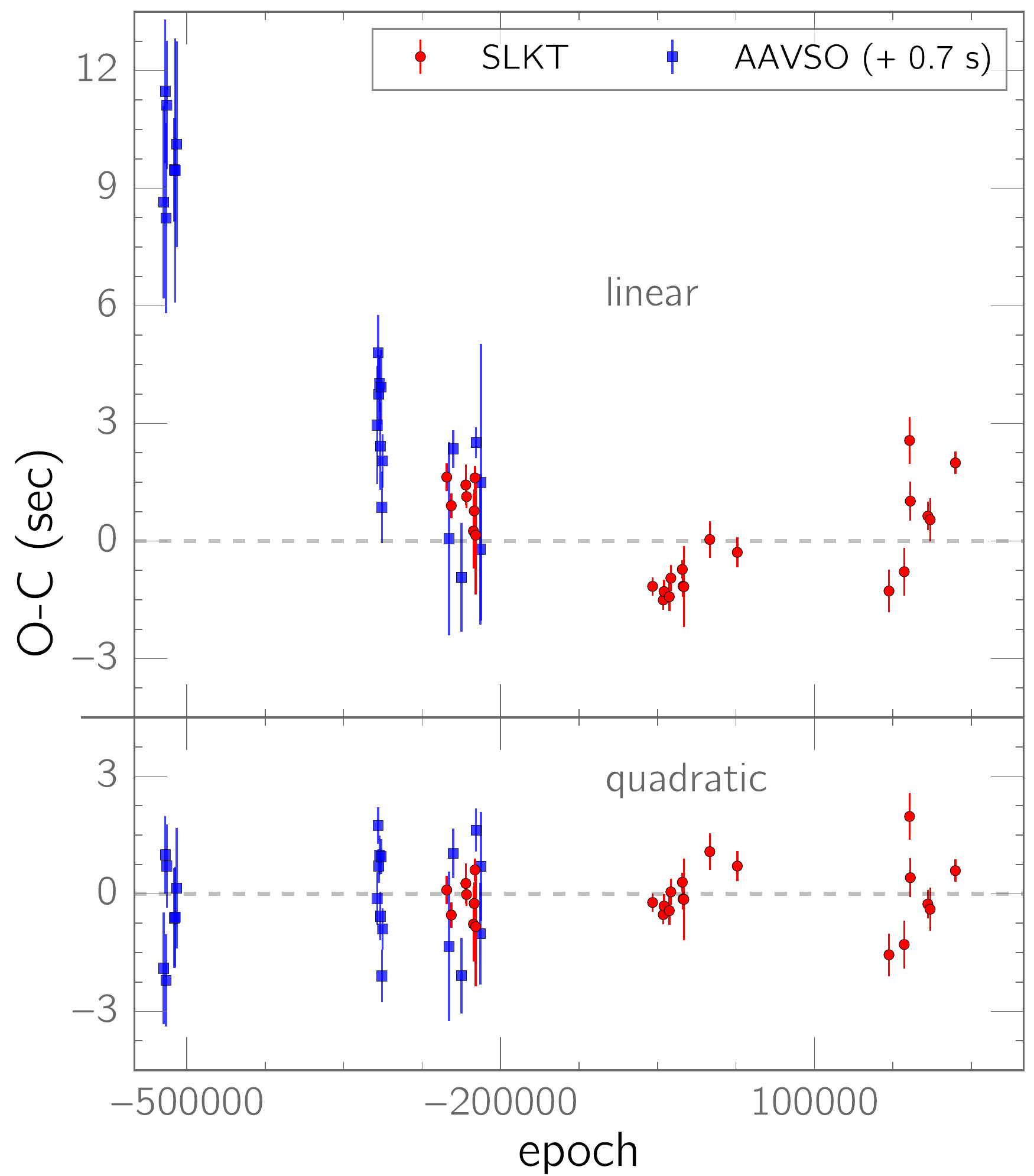}
\caption{A comparison of the residuals from the best-fit linear and quadratic ephemerides to beat-pulse timings from the SLKT (red markers) and AAVSO data (blue markers). Only the SLKT data were used to generate the ephemerides; the AAVSO data are shown as an independent test of both fits. Each SLKT point gives the mean and standard error of the residuals from one observing run. As described in the text, a constant offset of 0.7~sec was added to each AAVSO measurement to compensate for uncorrected shutter lag.}
\label{ephemeris_OC}
\end{figure}

We employed two independent fitting procedures---a bootstrap fit and an affine-invariant Markov Chain Monte Carlo algorithm \citep{emcee}---to generate linear and quadratic ephemerides for the pulse maxima. We inspected the residuals from the linear and quadratic ephemerides (Fig.~\ref{ephemeris_OC}) to determine which best described the data. While the measurements for each season systematically evade the linear fit, this trend vanished with the inclusion of a quadratic term, and the $\chi^{2}_{red}$ statistic dropped from 13.1 to 2.6. Based on the quadratic fit, we calculate a beat ephemeris of \begin{multline}
    T_{max}[BJD] = 4.91(31)\times10^{-16}E^2 +\\
    0.0013680458481(46)E + \\2457941.6688507(36). \label{ephem}
\end{multline} The quadratic coefficient is equivalent to $\frac{1}{2}\bar{P}_{beat}\dot{P}_{beat},$ where $\bar{P}_{beat}$ is the average beat period, yielding a unitless period derivative of $\dot{P}_{beat} = 7.18(45)\times10^{-13}.$ To convert this to a frequency derivative, we start with the definition $\nu = P^{-1}$ and differentiate with respect to $P$, obtaining $d\nu = -P^{-2}{dP}.$ The time derivative of this relation is \begin{equation} \dot{\nu} = -\frac{\dot{P}}{P^{2}}.\label{nu-dot}\end{equation} Thus, our $\dot{P}_{beat}$ is equivalent to $\dot{\nu}_{beat} = -5.14(32) \times 10^{-17}$ Hz s$^{-1}$.

\begin{figure}
    \centering
    \includegraphics[width=\columnwidth]{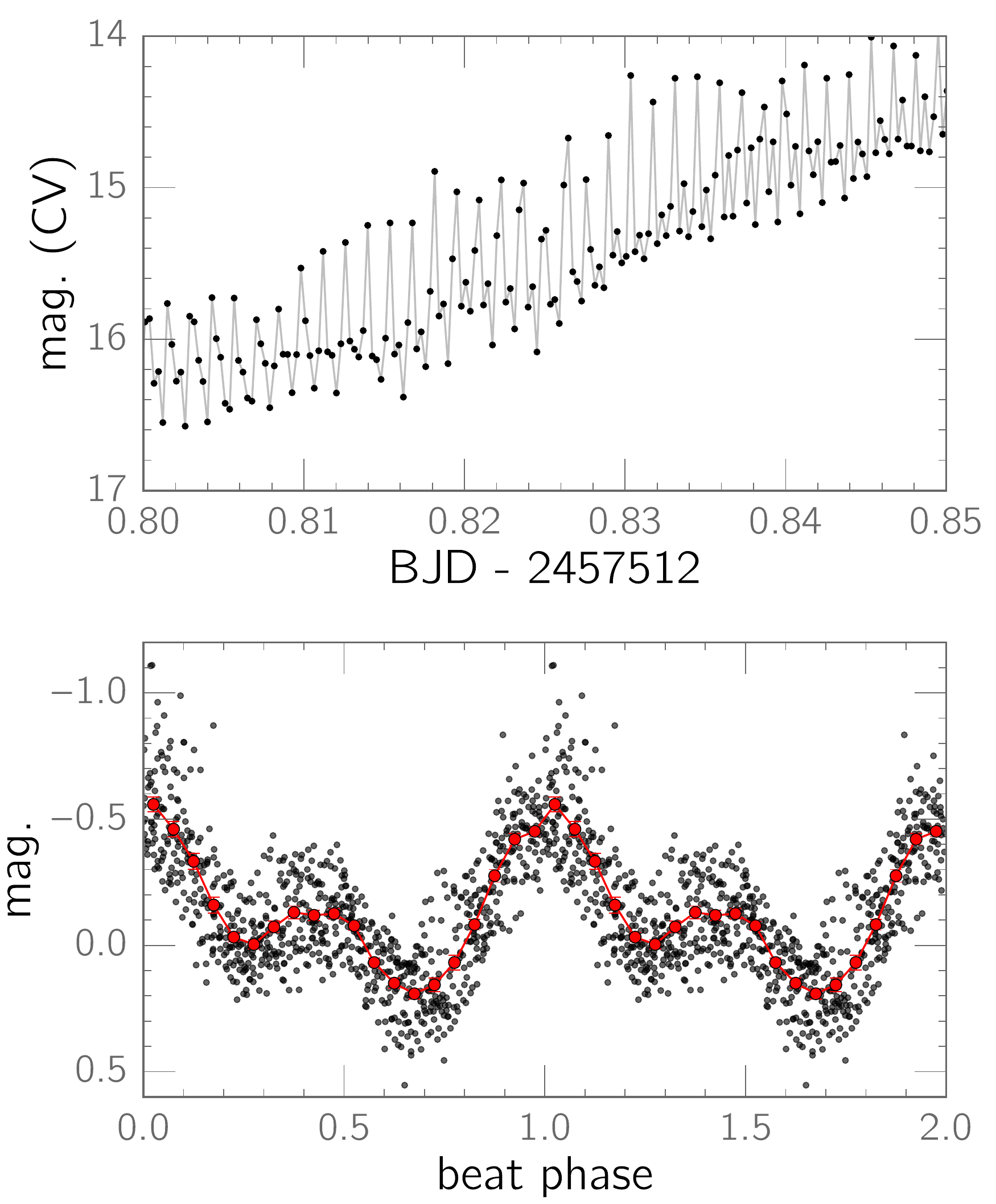}
    \caption{Top: a representative 1.2-hr section from a 5.7-hr AAVSO light curve whose image-to-image cadence was 24~s. Individual beat pulses are insufficiently resolved at this cadence for standard O$-$C analysis. The `CV' bandpass is unfiltered with a V zeropoint. Bottom: A phase plot of the beat pulse using the 
    the full 5.7-hr light curve. The waveform of the beat pulse is sufficiently well-sampled that its phase of maximum light can be measured. The red markers are phase bins.}
    \label{fig:aavso}
\end{figure}

\subsection{Testing the spin-down with AAVSO photometry}

Fig.~\ref{ephemeris_OC} includes O$-$C values from beat pulses in AAVSO photometry, but in order to provide an independent test of the spin-down ephemeris, these values were not included in the calculation of Eq.~\ref{ephem}. Because the cadences of these time series were too slow for O$-$C analysis of individual beat maxima, we took each AAVSO light curve and phased it to the beat period using our linear and quadratic ephemerides. The resulting beat-phase plots, an example of which is shown in Fig.~\ref{fig:aavso}, can be used to measure the average phase of the beat-pulse maxima provided that enough beat cycles were observed. We filtered the AAVSO photometry to exclude any time series with a cadence slower than 35~s or a duration shorter than 1~hr, and we applied a correction for the orbital-phase dependence of the pulse arrival times. We further excluded any time series afflicted by aliasing between the beat period and the observing cadence. 

With the benefit of the extended baseline of observations, it is immediately obvious in Fig.~\ref{ephemeris_OC} that the linear ephemeris leads to strong curvature in the residuals. By contrast, the quadratic residuals do not show any systematic trend.

No shutter-lag correction was applied to the AAVSO data, and the raw AAVSO quadratic residuals showed an offset of -0.7~s with respect to contemporaneous SLKT residuals. An uncorrected shutter lag will cause timestamps to be earlier than the data that they describe, imparting a small, negative O$-$C to the pulse timings. We therefore attribute the constant, negative offset of the AAVSO residuals to uncorrected shutter lag, and in Fig.~\ref{ephemeris_OC}, we add an offset of 0.7~s to each AAVSO residual to compensate for this effect. Regardless of the cause of this offset, the lack of a systematic trend in the AAVSO quadratic residuals supports our measurement of the spin-down.

\subsection{The nature of $\dot{\nu}_{beat}$}

\begin{figure}
    \centering
    \includegraphics[width=\columnwidth]{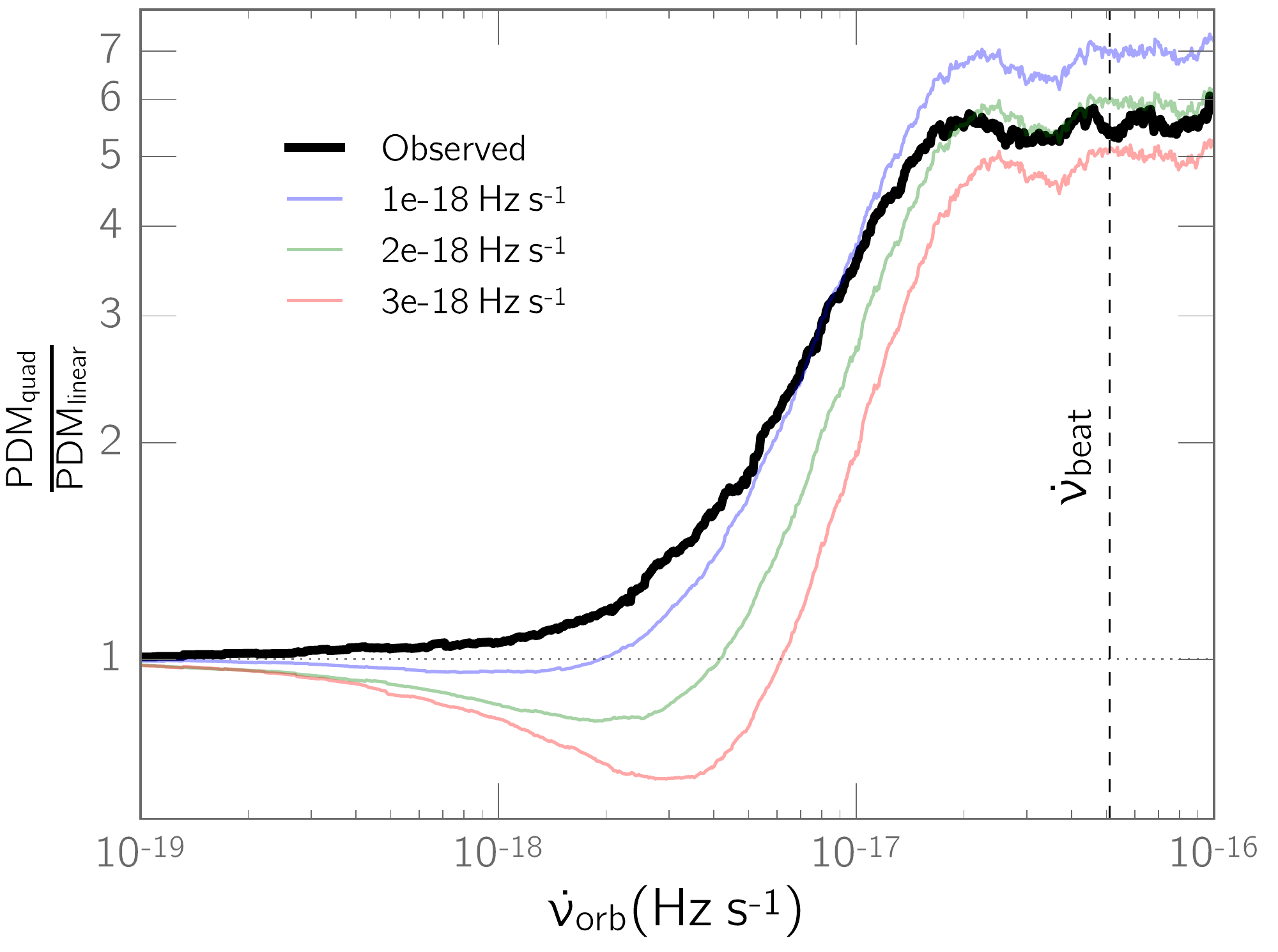}
    \caption{PDM analysis showing the effect of an increase in the orbital frequency across 13 years of CRTS and ASAS-SN photometry (thick black line). For each value of $\dot{\nu}_{orb}$, we calculated a quadratic orbital ephemeris, computed the PDM statistic of the resulting phase plot, and normalized it to the PDM statistic of the linear-ephemeris phase plot. A value greater than 1 (dotted horizontal line) indicates that adding $\dot{\nu}_{orb}$ increases the scatter in the phase plot, while values less than 1 indicate that $\dot{\nu}_{orb}$ reduces the scatter. The colored lines show the results of simulations in which an artificial $\dot{\nu}_{orb}$ was injected into a model of the orbital light curve. For $\dot{\nu}_{orb} \geq 2 \times 10^{-18}$ Hz s$^{-1}$, our technique successfully recovers the simulated $\dot{\nu}_{orb}$ at the global minimum of a given curve. We therefore conclude that $\dot{\nu}_{orb} < \sim 2 \times 10^{-18}$ Hz s$^{-1}$ and is too small to account for the observed $\dot{\nu}_{beat}$ (dashed vertical line). }
    \label{orbital_nu-dot}
\end{figure}


Because the beat frequency is defined as $\nu_{beat} = \nu_{spin} - \nu_{orb}$, its time derivative is $\dot{\nu}_{beat} = \dot{\nu}_{spin} - \dot{\nu}_{orb}$, meaning that the observed $\dot{\nu}_{beat}$ could be caused either by a decrease of the spin frequency or an increase in the orbital frequency. If the orbital period were changing so rapidly, there would be detectable consequences in long-term photometry. To explore this possibility, we used phase-dispersion minimization \citep[PDM; ][]{PDM} to determine whether an increasing orbital frequency could reduce the scatter in orbital phase plots using 13 years of survey photometry from the Catalina Real-Time Transient Survey \citep[CRTS;][]{drake} and All-Sky Automated Survey for Supernovae \citep[ASAS-SN;][]{shappee, kochanek}. For a range of $\dot{\nu}_{orb}$ between $1 \times 10^{-19}$ Hz s$^{-1}$ and $1 \times 10^{-16}$ Hz s$^{-1}$, we phased the survey photometry using a quadratic orbital ephemeris in which we assumed an initial orbital period of $P_0 = 0.14853528$~d \citep{marsh16} at an epoch of JD = 2455000.5 \citep{pb18}.\footnote{This is the epoch of the CRTS photometry used by \citet{marsh16} to measure the orbital period. It is different than the epoch in the orbital ephemeris in \citet{marsh16}, which reported a spectroscopically determined epoch and assumed---correctly---that the orbital period did not change appreciably in the intervening 5~years since the CRTS epoch. Phase plots using the CRTS epoch will have a uniform horizontal offset with respect to the \citet{marsh16} epoch, but this does not impact our PDM analysis.} For each quadratic phase plot, we computed the PDM statistic and normalized it to the the PDM statistic for the phase plot constructed with the linear ephemeris from \citet{marsh16}. The results of this analysis are shown in Fig.~\ref{orbital_nu-dot}. For $\dot{\nu}_{orb} > 2 \times 10^{-18}$ Hz s$^{-1}$, the phase plots contain an obvious increased scatter with respect to the linear ephemeris; below that value, the effect of $\dot{\nu}_{orb}$ becomes negligible, so we constrain $\dot{\nu}_{orb} < \sim2 \times 10^{-18}$ Hz s$^{-1}$. To verify this constraint, we created a simulated orbital light curve of AR Sco, injected an artificial $\dot{\nu}_{orb}$, and matched the sampling of the simulated data to the actual observations. As Fig.~\ref{orbital_nu-dot} shows, our algorithm successfully recovered values of $\dot{\nu}_{orb}$ in excess of $2 \times 10^{-18}$ Hz s$^{-1}$, but below that threshold, it could not discern the simulated $\dot{\nu}_{orb}$, bolstering our constraint.

The non-detection of $\dot{\nu}_{orb}$ is consistent with the measurements of the orbital phase of maximum light from \citet{littlefield17}. An increasing orbital frequency (\textit{i.e.,} $\dot{P}_{orb} < 0$) would induce concave-down curvature in their Fig.~4, but this is not seen. We conclude, therefore, that any $\dot{\nu}_{orb}$ contributes negligibly to $\dot{\nu}_{beat}$, such that $\dot{\nu}_{beat} = \dot{\nu}_{spin}$ to within our measurement uncertainty. Thus, the subscript notation becomes unnecessary for $\dot{\nu}$.

Because we have established that $\dot{\nu}$ is the same for the spin and beat frequencies, we can directly compare our measurement of $\dot{\nu}$ with its counterpart from \citet{marsh16}. Our estimate is larger by a factor of $\sim$1.8, but it still satisfies the constraint from \citet{pb18} that $\sim -2\times10^{-16}$ Hz s$^{-1} < \dot{\nu} <  \sim 1\times10^{-16}$ Hz s$^{-1}$.

\subsection{Reconciling \citet{marsh16} with \citet{pb18}}

\begin{figure}
    \centering
    \includegraphics[width=\columnwidth]{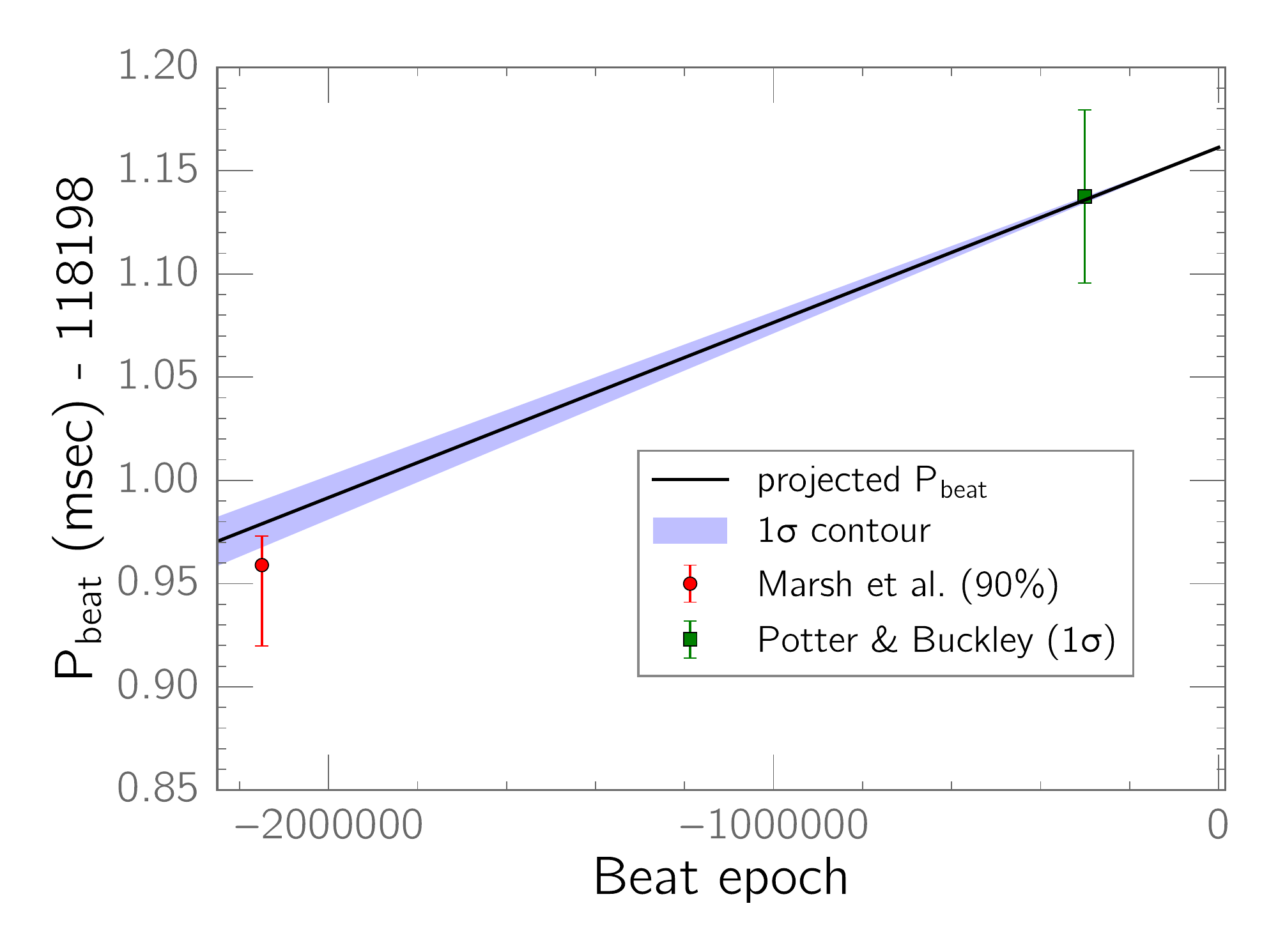}
    \caption{Projected beat period using Eq.~\ref{ephem}. The two markers indicate the beat periods reported by \citet{marsh16} and \citet{pb18}. Since \citet{pb18} did not explicitly provide a beat period, we calculated it from their spin period and the \citet{marsh16} orbital period. As indicated in the legend, \citet{marsh16} reported a 90\% confidence interval for their result, while \citet{pb18} provided a 1$\sigma$ interval. Even though these two points were not considered in the fitting procedure for Eq.~\ref{ephem}, the projected beat period is in excellent agreement with both measurements, substantiating our measurement of the WD's spin-down.}
    \label{fig:beat_period}
\end{figure}

The disagreement between the \citet{marsh16} spin-down ephemeris and the measured spin frequency in \citet{pb18} provides a stringent test of our spin-down ephemeris. In Fig.~\ref{fig:beat_period}, we plot the extrapolated beat period from Eq.~\ref{ephem} as a function of beat-cycle count, overlaying the beat periods from both \citet{marsh16} and \citet{pb18} at the appropriate epochs.\footnote{\citet{pb18} reported only $\nu_{spin}$, so we used $\nu_{orb}$ from \citet{marsh16} to calculate the corresponding $\nu_{beat}.$ } Even though our beat ephemeris was calculated without regard to either of these two measurements, the projected beat period is in excellent agreement with both measurements. Furthermore, the difference between the \citet{pb18} and \citet{marsh16} spin frequencies, when divided by the difference in epochs yields $\dot{\nu} = -(5.9\pm1.5)\times10^{-17}$ Hz s$^{-1}$, consistent with both our measurement ($\dot{\nu} = -5.14(32) \times 10^{-17}$ Hz s$^{-1}$) and the constraints from \citet{pb18}.

This suggests that the inability of the \citet{marsh16} ephemeris to correctly predict the spin and beat frequencies in \citet{pb18} is a consequence of an underestimated $\dot{\nu}$ and can be remedied by using our measurement of the spin-down. Despite the error in their estimate of $\dot{\nu}$, it is remarkable, given the sparse sampling and comparatively low time resolution of the survey photometry available to them, that they were able to accurately measure $\nu_{orb}$, $\nu_{beat}$, and $\nu_{spin}$ while also estimating $\dot{\nu}$ to within a factor of 2.

\subsection{Spin-down luminosity}

Our precise measurement of the frequency decay rate allows us to improve the estimate of the spin power available for conversion into the observed electromagnetic (EM) energy. The spin-down luminosity, is given by $L_{\dot{\nu}} =-4\pi^{2}I\nu_{spin}\dot{\nu},$ where $I$ is the WD's moment of inertia \citep{marsh16}. As did \citet{marsh16}, we assume a 0.8~M$_\odot$ WD with a radius of 0.01~R$_\odot$. The mass-radius relation for non-relativistic WD stars means that the moment of inertia changes as $I\propto M^{1/3}$, and is rather insensitive to variations in the assumed mass. In the non-relativistic regime, WDs are predicted to have a density structure like that of a polytrope with an index of 1.5, and we calculate a moment of inertia of $I = 0.25MR^2 = 2\times 10^{43}$ kg$\;$m$^2$. However, the strong magnetic fields and the rapid spin rate of the WD may have an effect on the precise value of the moment of inertia \citep{fs17}. 

Thus, we find the spin-down power is $3\times 10^{26}$~W. The Gaia DR2 parallax \citep{gaia16, gaia18} to AR~Sco ($8.492\pm0.041$~mas) is equivalent to a distance of $117.8\pm0.6$~pc, which is close enough to the \citet{marsh16} value that it is unnecessary to correct their measurement of the average EM power of the pulsations. We find the efficiency of converting the spin energy to detected EM emission to be $\sim$4\%.

\section{Conclusion}

The complex morphological variations seen in the optical light curve of AR Sco are explained as the superposition of the spin and beat pulses, as well as a slower orbital modulation. As noted in earlier studies, a beat cycle consists of a major and minor pulse. Here, we show the minor pulse has half the amplitude of the brighter signal after removing contamination from the spin pulse. We find the spin cycle also consists of two distinct pulses, but their amplitudes are comparable. The major spin pulse is half the amplitude of the major beat pulse and therefore, about the same amplitude as the minor beat pulse.

Over an orbit, the major beat and spin pulses add constructively between phases 0.2 and 0.3, resulting in the highest-amplitude optical variations observed in the system. Half an orbit later, the major beat and minor spin pulses add together along with the minor beat and major spin pulses. This combination results in lower amplitude peaks when compared with orbital phases around 0.25. The smallest amplitudes are seen when the beat and spin pulses are out of phase and destructively interfere around orbital phases between 0.5 and 0.6. This model provides a good overall fit to the rapid variations seen in AR~Sco. We also show that this model explains the O$-$C variations seen in the timings of the beat pulses. 

Our simple model assumes constant beat and spin amplitudes over an orbit, and this does not fully match the largest amplitudes fluctuations observed around orbital phases 0.2 to 0.4. A more complete model would take into account the changing viewing angle of the secondary star thought to be the origin of the beat emission. 

The major beat and spin pulses coincide at orbital phase 0.3, resulting in the highest amplitude peaks in the AR~Sco light curve. The beat pulse likely comes from near the surface of the red secondary star \citep{marsh16}, while the spin pulse is likely originating in the magnetic field of the WD. If we assume that beat maxima occur when one of the WD magnetic poles is pointing toward the secondary, then the geometry of the system suggests that we see the peak spin pulse when the WD magnetic pole is nearly perpendicular to our line-of-sight.

Perhaps most importantly, our results establish that the WD is indeed spinning down \citep[as originally proposed by][]{marsh16}, but the frequency derivative that we measure is almost twice as large as their estimate. Our updated spin-down ephemeris successfully passes two tests: it accurately predicts the evolution of the beat period between the \citet{marsh16} and \citet{pb18} epochs, and it also accounts for the pulse-arrival times in AAVSO photometry. Furthermore, it comports with the constraints on the spin-down rate established in \citet{pb18}. Our measurement of the spin-down of the WD confirms the conclusion by \citet{marsh16} that the pulsed EM emission from AR~Sco can be powered by the rotational energy of the WD.

\acknowledgments The Sarah L. Krizmanich Telescope is a generous donation by the Krizmanich family to the University of Notre Dame. It is named in honor of their daughter.

We thank David Buckley for showing us several figures from an early draft of the \citet{pb18} manuscript. We avoided using any of this advance information in our paper.

We are grateful to our anonymous referee for urging us to measure the spin-down rate of the white dwarf using the SLKT data.

This research uses observations of AR Sco from the AAVSO International Database contributed by GM and FJH, both of whom also participate in the Center for Backyard Astrophysics collaboration \citep{cba}.

\software{\\Astropy \citep{astropy}, emcee \citep{emcee}}
\facility{AAVSO}

\end{document}